\documentclass{jfm}
\usepackage{graphicx}
\usepackage{epstopdf, epsfig}

\shorttitle{Dynamics of bubble migration in a square channel flow of a viscoelastic fluid}
\shortauthor{Naseer et~al.}

\title{Dynamics of bubble migration in a square channel flow of a viscoelastic fluid}

\author{Hafiz Usman Naseer\aff{1},
  Daulet Izbassarov\aff{2},
  Zaheer Ahmed\aff{3}
 \and Metin Muradoglu\aff{1}
    \corresp{\email{mmuradoglu@ku.edu.tr}}}
    
\affiliation{\aff{1}Department of Mechanical Engineering, Koç University,
Istanbul, Türkiye
\aff{2}Finnish Meteorological Institute, Erik Palmenin aukio 1, 00560 Helsinki, Finland
\aff{3}Department of Mechanical Engineering, SZAB Campus, Mehran University of Engineering and Technology, Khairpur Mirs, Pakistan}

\begin{document}

\maketitle

\begin{abstract}
Cross-stream migration of a deformable bubble is investigated computationally in a pressure-driven channel flow of a viscoelastic fluid via interface-resolved simulations. The flow equations are solved fully coupled with the Giesekus model equations using the front-tracking method and extensive simulations are performed for a wide range of flow parameters to reveal the effects of bubble deformability, fluid elasticity, shear-thinning, and fluid inertia on the bubble migration dynamics. Migration rate of a bubble is found to be much higher than that of a solid particle under similar flow conditions mainly due to free-slip condition on its surface. It is observed that direction of bubble migration can be altered by varying shear-thinning of the ambient fluid. With a strong shear-thinning, the bubble migrates towards the wall while it migrates towards the center of the channel in a purely elastic fluid without shear-thinning. An onset of elastic flow instability is observed beyond a critical Weissenberg number, which in turn causes a path instability even for a nearly spherical bubble. An inertial path instability is also observed once bubble deformation exceeds a critical value. Shear-thinning is found to be suppressing the path instability in a viscoelastic fluid with a high polymer concentration whereas it reverses its role and promotes path instability in a dilute polymer solution. It is found that bubble migration towards wall induces a secondary flow with a velocity that is about an order of magnitude higher than the one induced by a solid particle under similar flow conditions.
\end{abstract}


\section{Introduction}
Cross-stream or lateral migration of particles in a pressure-driven flow has been an active area of research due to its relevance in many industrial (\cite{shannon2008science}, \cite{vidic2013impact}) and biological (\cite{nagrath2007isolation}) applications. The phenomenon of lateral migration is specifically used for the manipulation of biological cells and particles in various microfluidic applications (\cite{xuan2010particle}, \cite{amini2014inertial}). The wall normal component of force acting on the particle is responsible for its lateral migration and it is usually called lift force. The lift force which is often relatively small in magnitude plays a central role in determining the final position of an individual particle and the particle distribution in particulate flows (\cite{hidman2022lift}). 

The lateral motion of a particle was first investigated experimentally by \cite{segre1961radial} in a tube flow. Their observation was later confirmed by many experimental (\cite{karnis1966particle}, \cite{matas2004inertial}) and numerical works (\cite{feng1994direct}, \cite{yang2005migration}). The same observation was made in rectangular channels where particles move towards or away from the channel wall in the cross-streamwise direction (\cite{frank2003particle}, \cite{del2013particle}, \cite{wang2018numerical}, \cite{yuan2018recent}). The situation becomes more involved when the ambient fluid is viscoelastic as the fluid elasticity plays its role in pushing the particle towards the center of the channel (\cite{seo2014lateral}). In the case of viscoelastic fluids, most of the existing literature is primarily focused on solid particles suspended in a viscoelastic fluid (\citet{villone2013particle}). In the absence of inertia, the cross-stream migration of a non-deformable particle is not possible to occur in a confined Newtonian fluid flow due to the inherent reversibility of Stokes flow. On the other hand, the non-linear characteristics of viscoelastic constitutive equations provide the required irreversibility making the lateral migration possible in a complex fluid even with negligible inertia. The lateral migration of particles in complex fluids and their application in manipulating the particles in various microfluidic devices have been summarized in the review paper by \citet{d2017particle}.

In the presence of both inertia and viscoelasticity, several factors determine the orientation of particle migration in a pressure-driven channel flow. Some of these factors reported in numerous studies so far include shear-induced lift force (\citet{saffman1965lift}), wall-induced lift force (\citet{article}), fluid elasticity (\citet{karnis1966particle}), initial position of the particle (e.g.,\citet{villone2011simulations}), geometry of the channel, the shear-thinning of the ambient fluid (\cite{li2015dynamics}) and particle rotation rate (the Magnus effect). The effects of one or more of these factors determine  the migration of a particle towards or away from the channel wall. Moreover, the deformability of a bubble adds further complexity to this phenomenon, and therefore, complex dynamics of lateral migration of bubbles in viscoelastic fluids are yet to be fully explored. As a result of this migration, some interesting secondary flow features also emerge in the vicinity of the particle. One of them is a secondary flow which is perpendicular to the primary flow (streamwise) direction and may have a velocity as large as the order of the particle migration velocity. However, this secondary flow is generated in non-circular channels only by the viscoelastic fluids having non-zero second normal stress difference (\citet{debbaut1997development}).

When a spherical particle moves with a relative velocity in a shear flow, a force is exerted on the particle by the surrounding fluid in a direction perpendicular to its relative motion. This force is known as the shear-induced lift force, first calculated analytically by \citet{saffman1965lift} for a solid sphere. This force pushes the particle towards the channel wall until it is balanced by the wall-induced lift force (\cite{feng1994direct}, \cite{matas2004lateral}). The flow field around the particle is significantly influenced by the presence of the wall. The wall decelerates the motion of the particle in the streamwise direction due to extra drag and repels it away from the wall if the characteristic length of the particle is much smaller than the channel size. If the characteristic length of the particle is comparable to the channel size, like the motion of bubbles in capillaries, the channel walls constrain the motion of the immersed object, as explained by \citet{michaelides2006particles}.

The well-known Magnus effect (rotation-induced lift force) is another factor influencing the migration of an object in the channel flow. For a solid particle having the no-slip condition on its surface, the difference in the relative fluid velocity on its sides may cause it to rotate. This shear-induced rotation generates an additional transverse pressure difference and the resulting lift force is referred to as the `Magnus effect' (\cite{rubinow1961transverse}). Instead of a solid particle, if the object is deformable like a bubble, the deformability induces yet another lateral migration directed toward the channel center (\cite{chan1979motion}). Similarly, the initial position of the particle in the channel has a direct impact on its lateral migration as dictated by the gradient of velocity at that initial location. The size of the particle as compared to the channel is another important variable to be considered while analyzing the lateral migration in a pressure-driven flow. It is generally characterized by the blockage ratio (\cite{di2009inertial}) defined as $b=d/H$, where $d$ is  the particle  size (e.g., diameter) and $H$ is the channel width. It has been observed in both experimental (\cite{lim2014inertio}) and numerical (\cite{mortazavi2000numerical}) works that a bubble with a diameter beyond a certain limit migrates towards the channel center due to deformation-induced lift force while a smaller one migrates towards the wall due to inertial lift force. This feature is used in many microfluidic devices for sorting particles of different sizes (\cite{nam2015microfluidic}). 

The Magnus effect, shear-induced lift force, and wall-induced lift force may be referred to as the `inertial effects' as all these forces primarily originate due to fluid inertia. If the ambient fluid is viscoelastic, the elastic effects tend to push the object toward the center (\cite{seo2014lateral}) while the inertial effects tend to move it toward the channel wall. The relative strength of inertia and elasticity determines the final equilibrium position of a non-deformable particle in a channel. This interplay between inertia and elasticity is quantified by the elasticity number, defined as $El$= $Wi/Re$, where $Wi$ and $Re$ are the Weissenberg number and the Reynolds number, respectively. Furthermore, if the ambient fluid exhibits a shear-thinning behavior as well, it has been reported by \cite{li2015dynamics} that the shear-thinning amplifies inertial effects and thus promotes particle migration towards the wall. 

The above-mentioned list of factors affecting the dynamics of bubble migration is still not exhaustive. In the presence of surfactant contamination, the surfactant-induced Marangoni stresses oppose the inertial lift force and can completely alter the dynamics of bubble migration. In our earlier work (\citet{muradoglu2014front,ahmed2020effects}), it was observed that, in the pressure-driven channel flow of a Newtonian fluid, clean spherical bubbles move towards the wall while the deformable ones migrate away from it. On the other hand, even the spherical bubbles can move away from the wall in the presence of a strong enough surfactant. 

Although there are a large number of experimental and numerical works devoted to solid particle migration, much less attention has been paid to the combined effects of inertia, viscoelasticity, and shear-thinning on the lateral migration of a deformable bubble and only a few details are available in the existing literature, which motivates the present study. The main focus here is to explore the combined effects of fluid inertia, viscoelasticity, and bubble deformability on its lateral migration in a shear-thinning viscoelastic fluid. For this purpose, fully interface-resolved numerical simulations are performed using a finite-difference/front-tracking method for a wide range of relevant flow parameters to explore the features of this complex flow field. In addition to the lateral migration, the bubble-induced elastic instability and its impact on the onset of a bubble path instability,  the effect of shear-thinning, and the resulting flow  are investigated computationally. The secondary flow field developed due to the second normal stress difference in the flow is also examined and quantified.  

The rest of the paper is organized as follows. The governing equations and numerical methods are briefly described in the next section. The problem statement and the computational setup are described in $\S$\ref{CS}. The results are presented and discussed in $\S$\ref{Results} followed by the conclusions in $\S$\ref{Conclusion}. A grid convergence study is conducted and the results are presented in the Appendix.

\section{Governing Equations and Numerical Method}\label{NM}

The flow equations and the Giesekus model are described here in the context of the finite-difference/front-tracking method that utilizes a one-field formulation. As discussed by \cite{tryggvason2011direct,ahmed2020turbulent,izbassarov2021polymer}, \cite{izbassarov2015front}, one set of governing equations can be written for the entire multiphase computational domain. In this approach, the effect of surface tension is taken into account by adding a distributed body force term to the momentum equations near the interface, and the discontinuities in material properties are handled by defining an indicator function. The Navier-Stokes equations are thus written as

\begin{eqnarray}
\rho\frac{\partial \mathbf{u}}{\partial t}
+{\rho} \boldsymbol{\nabla}\cdot({\mathbf{u}}{\mathbf{u}})
=-\boldsymbol{\nabla}{p} -\frac{dp_0}{dy}\mathbf{j}
+\boldsymbol{\nabla}\cdot{\boldsymbol\tau} 
+\boldsymbol{\nabla}\cdot\mu_s(\boldsymbol{\nabla}{{\mathbf{u}}}+\boldsymbol{\nabla}{{\mathbf{u}}^T})
+\mathbf{g}\left(\rho-\rho_{ave}\right)  \nonumber \\
+ \int_A \sigma\kappa\mathbf{n}\delta(\mathbf{x}-\mathbf{x}_f)dA,
\label{NS}
\end{eqnarray}

\noindent 
where ${\mathbf{u}}$, $\boldsymbol{\tau}$, $p$, $\rho$, $\mu_s$ are the velocity vector, polymer stress tensor, pressure, discontinuous density, and solvent viscosity fields, respectively. A constant pressure gradient $-\frac{dp_0}{dy}\mathbf{j}$ is applied to drive the flow where $\mathbf{j}$ is the unit vector in the $y$-direction. The buoyancy term consists of the gravitational acceleration $\mathbf{g}$ and the density difference $\rho-\rho_{ave}$, where $\rho_{ave}$ is the average density in the computational domain. The effect of surface tension is added as a body force term on the right-hand side of the momentum equation where $\sigma$ is the surface tension coefficient, $\kappa$ is twice the mean curvature and $\mathbf{n}$ is a unit vector normal to the interface. As the surface tension acts only on the interface, $\delta$ represents a three-dimensional Dirac delta function with the arguments $\mathbf{x}$ and $\mathbf{x}_f$ being a point at which the equation is evaluated and a point at the interface, respectively. The momentum equation is supplemented by the incompressibility condition
 \begin{equation}
 \boldsymbol{\nabla} \cdot \mathbf{u} = 0.
 \label{continuity}
 \end{equation}

\noindent It is assumed that the material properties remain constant following a fluid particle, i.e.,
\begin{equation}
\frac{D\rho}{Dt} =0,  \frac{D\mu_s}{Dt} =0,  \frac{D\mu_p}{Dt} =0,  \frac{D\lambda}{Dt} =0,
\end{equation}

\noindent where $\frac{D}{Dt}= \frac{\partial}{\partial t} + \mathbf{u}\cdot\nabla$ is the material derivative. For a viscoelastic fluid, $\mu_p$ is the polymeric viscosity and $\lambda$ is the polymer relaxation time. 

Using an indicator function ($I$), the material properties are set in the entire computational domain as
\begin{equation}
\left. \begin{array}{llll} 
\rho =\rho_{o}I({\mathbf{x}},t)+\rho_i\left(1-I({\mathbf{x}},t)\right),
\quad \\[8pt]
\mu_s =\mu_{s,o}I({\mathbf{x}},t)+\mu_{s,i}\left(1-I({\mathbf{x}},t)\right),
\quad \\[8pt]
\mu_p =\mu_{p,o}I({\mathbf{x}},t)+\mu_{p,i}\left(1-I({\mathbf{x}},t)\right),
\quad \\[8pt]
\lambda =\lambda_{o}I({\mathbf{x}},t)+\lambda_i\left(1-I({\mathbf{x}},t)\right),
\end{array}\right\}
\label{IndicatorFunction1}
\end{equation}

\noindent where the subscripts ``$i$" and ``$o$" denote the properties of the bubble and the bulk fluid, respectively. The indicator function is defined as

\begin{equation}
  I({\mathbf{x}},t) = \left\{
    \begin{array}{ll}
      1 & \mbox{in bulk fluid,\ } \\[2pt]
      0 & \mbox{in bubble domain.\ }
    \end{array} \right.
\end{equation}

Viscoelasticity of bulk liquid is modeled using the Giesekus model (\cite{giesekus1982simple}). This model is capable of capturing the elongation of individual polymer chains and the resulting shear-thinning behavior of the viscoelastic fluid. In the Giesekus model, the polymer stress tensor $\boldsymbol{\tau}$ evolves by

\begin{equation}
\boldsymbol{\tau} = \frac{\mu_p}{\lambda}(\boldsymbol {B} - \boldsymbol{I}), 
\label{stress_eq}
\end{equation}
where $\boldsymbol{B}$ and $\boldsymbol{I}$ are the conformation and the identity tensors, respectively. The conformation tensor evolves by

\begin{equation}
 \frac{\partial \boldsymbol{B}}{\partial t} + \mathbf{u}\cdot\boldsymbol{\nabla}\boldsymbol{B} - \boldsymbol{\nabla}\mathbf{u}^T \cdot \boldsymbol{B} - \boldsymbol{B}\cdot\boldsymbol{\nabla}\mathbf{u} = \frac{1}{\lambda} [(1-\alpha)\boldsymbol{I} + (2\alpha - 1)\boldsymbol{B} - \alpha\boldsymbol{B^2}],
\label{conformation_tensor}
\end{equation}


\noindent where $\alpha$ is the mobility factor representing the anisotropy of the hydrodynamic drag exerted on the polymer molecules. Due to the thermodynamic considerations, $\alpha$ is restricted to $0\le \alpha\le 0.5$ (\cite{schleiniger1991remark}). When $\alpha = 0$, the Giesekus model reduces to the Oldroyd-B model. 

At high Weissenberg numbers, these highly non-linear viscoelastic constitutive equations become extremely stiff, which makes their numerical solution a challenging task. The problem is overcome by utilizing the well-known log-conformation method where an eigen decomposition is employed to re-write the constitutive equation of the conformation tensor in terms of its logarithm (\cite{izbassarov2015front,izbassarov2018computational}). The interested readers are referred to \cite{fattal2005time} for the detailed procedure.

The flow equations (Eqs.~(\ref{NS}) and (\ref{continuity})) are solved fully coupled with the Giesekus model equation (Eq.~(\ref{stress_eq})). A QUICK scheme is used to discretize the convective terms in the momentum equations while second-order central differences are used for the diffusive terms. For the convective terms in the viscoelastic equations, a 5th-order WENO-Z (\cite{borges2008wai}) scheme is used. An FFT-based solver is used for the pressure Poisson equation. Since the pressure equation is not separable due to variable density in the present multiphase flow, the FFT-based solvers cannot be used directly. To overcome this challenge, a pressure-splitting technique presented by \cite{dong2012time} and \cite{dodd2014fast} is employed. A predictor-corrector scheme is used to achieve second-order time accuracy as described by \citet{tryggvason2001front}. The details of the front-tracking method can be found in the book by \cite{tryggvason2011direct} and in the review paper by \cite{tryggvason2001front}, and the treatment of the viscoelastic model equations in  \citet{izbassarov2015front} and \citet{izbassarov2018computational}.

\section{Problem Statement and Computational Setup}\label{CS}

\begin{figure}
  \includegraphics[width=1.0\textwidth]{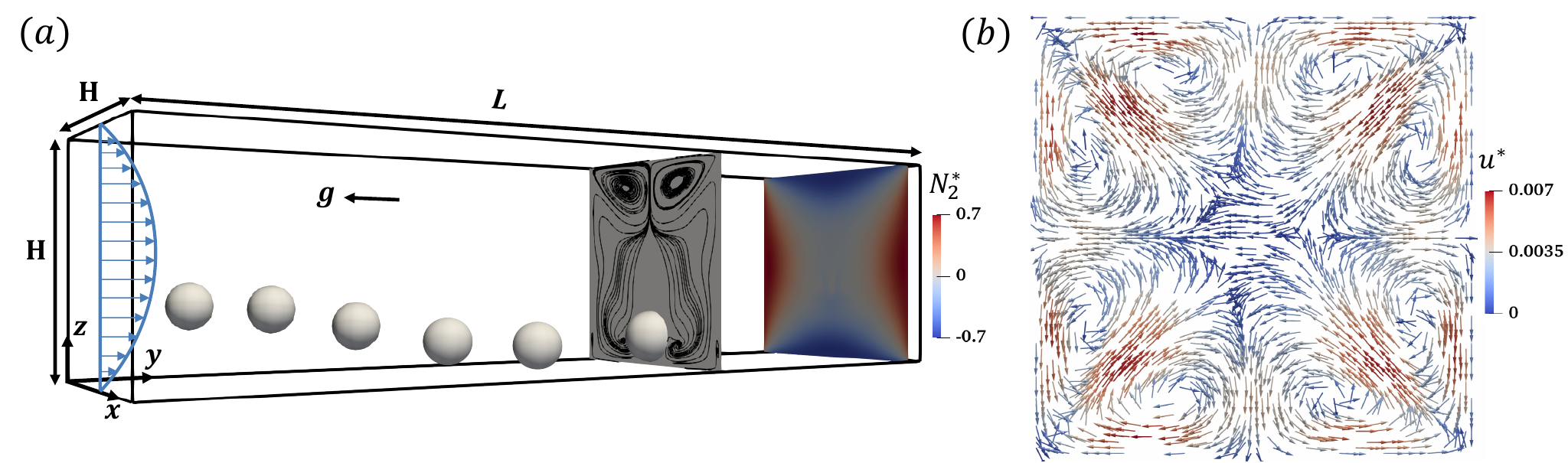}
  \caption{(\textit{a}) The computational domain is shown with the schematic representation of a bubble migrating toward the wall of the channel. The streamlines of the flow are shown above the bubble in the $xz$-plane once the bubble reaches its equilibrium position closer to the wall. The distribution of the second normal stress difference ($N_2$) is shown in the $xz$-plane upstream of the bubble at $y=3.95$. (\textit{b}) Eight vortices showing the secondary flow field in the same $xz$-plane away from the bubble at $y=3.95$ are shown. The velocity vectors are colored by the magnitude of flow velocity in that plane. ($Re=18.9, El=0.05, Eo=0, Ca=0.01, \alpha=0.1$)}
\label{domain}
\end{figure}

Microfluidic devices usually use channels with rectangular cross-sections due to their relative ease of fabrication. If the aspect ratio of the channel is reduced, the relative influence of channel walls on the flow field becomes more pronounced (\citet{villone2013particle}). The limiting case is obtained for a square-shaped cross-section, which is selected in the present study. Figure~\ref{domain}a shows the computational domain which is a square channel with the dimensions of $H$, $L$, and $H$ in the $x$, $y$, and $z$ directions, respectively. Periodic boundary conditions are applied in the streamwise ($y$) direction whereas the other two directions ($x$ and $z$) have no-slip boundary conditions. The centroid of the bubble is denoted by $(x_b,y_b,z_b)$ and a spherical bubble of diameter $d$ is initially located at $(x_b,y_b,z_b) = (0.5H, H, 0.25H)$. The value of $H$ is set to $4d$. After performing the simulations by gradually increasing the length of the channel in the streamwise ($y$) direction to check the effects of periodicity, the channel length is set to $L=4H$, which is found to be sufficient to eliminate any significant effects of periodicity. 

The bubble is at rest initially at $t=0$ and a constant pressure gradient $dp_0/dy = -U_{o}\mu_{o}\pi^3 / 4kH^2$ is applied in the $y$-direction to drive the flow, where $U_{o}$ is the flow velocity at the centerline of the channel in the case of a fully-developed single-phase laminar flow and $k$ is a geometric constant. For a square channel, $k\approx0.571$ (\citet{fetecau2005unsteady}). It is important to note that $U_o$ is the centerline velocity in the channel for the Newtonian and Oldroyd-B fluids. In a Giesekus fluid, the centerline velocity becomes greater than $U_o$ due to the shear-thinning effect. However, throughout this paper, all the non-dimensional numbers are defined based on $U_o$ corresponding to the applied pressure gradient in the Newtonian fluid unless stated otherwise. $H$, $U_{o}$, and $H/U_{o}$ are used as the length, velocity, and time scales, respectively. The stresses are normalized by $\mu_{o}U_{o}/H$. The normalized non-dimensional quantities are denoted by the superscript ($^*$). The blockage ratio ($b=d/H$) is fixed at $0.25$. The density ($\rho_{o}/\rho_{i}$) and the viscosity ($\mu_{o}/\mu_{i}$) ratios are set to $10$ in this study. These comparatively smaller ratios are used to enhance numerical stability and thus relax the time step restrictions. \citet{tasoglu2010impact} and \citet{olgac2013direct} have demonstrated that the results are not affected significantly for these types of flows at higher ratios. In the present case of pressure-driven viscoelastic channel flow, although not shown here, simulations have been performed for the density and viscosity ratios of $20$ and $40$ as well, and the results are found to be essentially the same as those computed for $\rho_{o}/\rho_{i} = \mu_{o}/\mu_{i} = 10$, i.e., the difference is less than 1\%.

The flow conditions are characterized by the following non-dimensional numbers defined as

\begin{eqnarray}
Re = \frac{\rho_o U_o H}{\mu_o},
Wi = \frac{\lambda U_o}{H},
Eo = \frac{g\Delta\rho d^2}{\sigma},
Mo = \frac{g\mu_o^4}{\rho_o\sigma^3},
Ca = \frac{\mu_o U_o}{\sigma},
\beta = \frac{\mu_s}{\mu_o},
\label{ND_numbers}
\end{eqnarray} 

\noindent where $Re$, $Wi$, $Eo$, $Mo$, and $Ca$ denote the Reynolds, Weissenberg, E{\"o}tv{\"o}s, Morton, and capillary numbers, respectively. $\beta$ is the ratio of solvent viscosity to zero shear viscosity of the viscoelastic fluid. The density difference is defined as $\Delta\rho = \rho_o-\rho_i$. The bubble deformation ($\chi$) is quantified as
\begin{equation}
 \chi = \sqrt{I_{\rm max}/I_{\rm min}} \ ,
 \label{deformation}
\end{equation}

\noindent where $I_{\rm max}$ and $I_{\rm min}$ are the maximum and minimum eigenvalues of the second moment of inertia tensor defined as
\begin{equation}
  I_{ij} = \frac{1}{\mathcal{V}_b}\int_{V}(x_{i} - x_{io})(x_{j} - x_{jo})dV,
 \label{momentofinertia}
\end{equation}

\noindent where $\mathcal{V}_b$ is the volume of the bubble and $x_{io}$ and $x_{jo}$ are the coordinates of the bubble centroid in the $i^{th}$ and $j^{th}$ directions, respectively. \cite{bunner2003effect} has shown that deformation quantified by this method is approximately equal to the ratio of the shortest to the longest axis for modestly deformed ellipsoids. However, for complex bubble shapes, this definition of deformation gives a more general measure for the bubble deformation and eliminates uncertainty in identifying the longest and the shortest axes.
\section{Results and Discussions}\label{Results}
We first examine the dynamics of a nearly spherical bubble under similar conditions as used by \citet{li2015dynamics} for a solid particle, i.e., $Re=18.9, El=0.05, Ca=0.01, \alpha=0$. The main difference in the present case is the slip at the bubble interface. Note that this set of parameters is also designated as the {\it baseline} case in the present study to facilitate direct comparison with the results obtained for a solid particle by \citet{li2015dynamics}. 

A peculiar feature of a viscoelastic fluid flow is the presence of normal stress differences, which are given by $N_1 = \tau_{yy} - \tau_{zz}$ and $N_2 = \tau_{zz} - \tau_{xx}$ in the present scenario. In the absence of the shear-thinning effect ($\alpha=0$), the Giesekus model reduces to the Oldroyd-B model and the second normal stress difference ($N_2$) becomes zero. Thus, $\alpha>0$ is an essential condition to model a viscoelastic fluid that exhibits a non-zero second normal stress difference. The geometry of the channel also plays a significant role in the development of these stresses. $N_1$ is found to be maximum in the highest shear region near the walls and its value becomes minimum in the corners and at the center. \citet{ho1976migration} argued that this particular distribution of $N_1$ in a four-wall channel is the primary reason for the accumulation of solid particles in the corner and centerline regions. When the second normal stress difference develops in a shear-thinning viscoelastic fluid, not only does the distribution of $N_1$ in the channel change but its magnitude is also reduced due to enhanced inertial effects (\citet{li2015dynamics}). Figure~\ref{domain}a shows the distribution of $N_2$ in the channel cross-section when there is a strong shear-thinning effect present in the flow, i.e., $\alpha=0.1$. Eight vortices generated due to this distribution of $N_2$ away from the bubble are also depicted in Fig.~\ref{domain}b. This particular distribution of $N_2$ affects the orientation of bubble migration and the development of a secondary flow field around the bubble. An asymmetric pattern of streamlines is shown in Fig.~\ref{domain}a above the bubble once the bubble reaches its equilibrium position near the wall. These effects on the lateral migration of a deformable bubble are explored one by one in the subsequent sections. 
\subsection{Dynamics of Bubble Migration}\label{4.1}
\begin{figure}
  \includegraphics[width=1.0\textwidth]{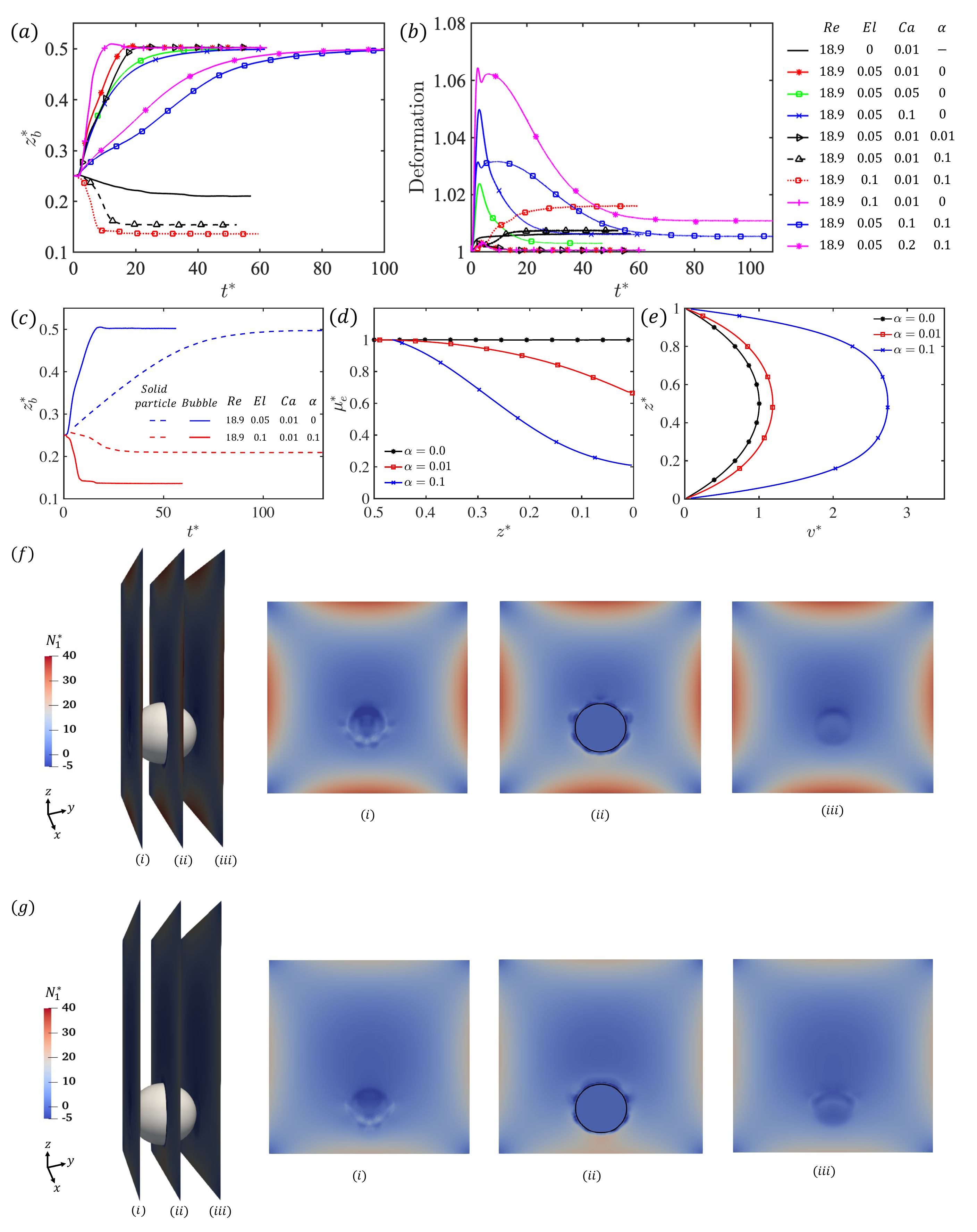}
  \caption{(\textit{a}) Evolution of bubble displacement in the wall-normal ($z$) direction under non-buoyant conditions. (\textit{b}) The corresponding change in the bubble deformation is shown as the bubble moves towards or away from the wall. (\textit{c}) The evolution of bubble displacement is compared with a solid particle studied by \citet{li2015dynamics} under similar flow conditions. (\textit{d}) The reduction in the effective viscosity of the fluid due to shear-thinning  is quantified in a vertical cutting $xz$-plane away from the bubble. (\textit{e}) The velocity profile in the same $xz$-plane is shown for different values of $\alpha$. The distributions of the first normal stress difference ($N_1$) around the bubble are shown in  vertical cutting planes of $(i), (ii)$, and $(iii)$ as indicated on the left for (\textit{f}) $\alpha=0.01$ and for (\textit{g}) $\alpha=0.1$. For (\textit{d}, \textit{e}, \textit{f} \& \textit{g}) $Re=18.9, El=0.05, Ca=0.01$.} 
\label{displacement}
\end{figure}

Simulations are first performed to examine the effects of the elasticity number, the capillary number, and the mobility factor on the lateral migration of a non-buoyant bubble in the channel flow at the nominal Reynolds number $Re = 18.9$. Figures~\ref{displacement}a and \ref{displacement}b show the evolution of bubble displacement in the lateral ($z$) direction and the corresponding change in the bubble deformation during its migration, respectively. In a Newtonian fluid, a nearly spherical bubble slightly moves towards the channel wall under the effects of inertia and stabilizes near the wall where wall-induced repulsive force balances the lift force. When the ambient fluid is viscoelastic, modeled by Oldroyd-B ($\alpha = 0$), the same spherical bubble moves towards the center of the channel under the elastic effects just like a solid particle (\cite{li2015dynamics}). It is observed that, although the overall trend of bubble displacement in the Oldroyd-B fluid is similar to that of a solid particle, the bubble moves towards the center of the channel at a much higher rate as shown in Fig. \ref{displacement}c. Compared to a solid particle, a bubble experiences a smaller drag due to  slip at the interface, which makes it more sensitive to the inertial and elastic effects. When the bubble deformability is increased gradually by increasing the capillary number to $Ca = 0.05$ and $Ca = 0.1$ for the same Oldroyd-B fluid, the rate of bubble migration towards the channel center is slightly reduced. As the bubble moves towards the low shear region (towards the center), its deformability starts to reduce and the same effect is observed in its lateral velocity as well. The effects on bubble velocity will be discussed in detail later in Section \ref{4.5}. It is interesting to observe that, although the deformability-induced lift force is expected to push the bubble towards the center, it acts in the opposite direction at this low Reynolds number and slows down the bubble migration towards the center of the channel. In fact, the deformability-induced lift force may change its sign depending upon the flow conditions as previously observed both experimentally (\citet{tomiyama2002transverse}) and numerically (\citet{ervin1997rise}). The governing mechanisms for the deformability-induced lift force and the change in its sign (direction) have been recently examined in detail by \citet{hidman2022lift}.

When the shear-thinning effects are enhanced by increasing the mobility parameter ($\alpha$) in the Giesekus model, the orientation of bubble migration changes due to an increase in the relative importance of the fluid inertia. At a very small value of $\alpha=0.01$, a spherical bubble still moves towards the center but it takes a comparatively longer time than that in the Oldroyd-B fluid to reach its equilibrium position. Once $\alpha$ is increased further to $0.1$, the orientation of bubble migration changes completely, i.e., instead of moving towards the center, it starts moving towards the wall. Again, although the trend is similar to that of a solid particle in a shear-thinning fluid under the same parametric settings, the bubble migration is significantly more sensitive to the shear-thinning parameter than the corresponding solid particle. For example, while $\alpha = 0.2$ is required by the solid particle to reverse its orientation from the channel center towards the wall (\cite{li2015dynamics}), $\alpha=0.1$ is sufficient for the bubble to reverse its orientation. The shear-thinning effect is quantified by the reduction in the effective viscosity ($\mu_{e}$) of the fluid defined as 

\begin{equation}
 \mu_e/\mu_o = \frac{\mu_s(\frac{\partial v}{\partial z}+\frac{\partial w}{\partial y})+\tau_{yz}}{\mu_o(\frac{\partial v}{\partial z}+\frac{\partial w}{\partial y})} .
 \label{mu_e}
\end{equation} 

\noindent The variation of the effective viscosity and the corresponding flow velocity profiles are shown Figs.~\ref{displacement}d and \ref{displacement}e, respectively, for $\alpha =0$, 0.01 and 0.1. As seen, for $\alpha = 0$, the effective viscosity remains constant in the channel as expected. In a shear-thinning fluid ($\alpha>0$), the viscosity of the fluid decreases significantly in the high shear region near the wall and becomes as low as $20\%$ of its value in the channel center for $\alpha=0.1$. As a result, the centerline flow velocity becomes approximately $2.8$ times larger than that of the non-shear-thinning fluid (Fig.~\ref{displacement}e). This enhanced inertia helps the quick migration of the bubble towards the wall as seen in Fig.~\ref{displacement}a. The shear-thinning also changes the distribution of the first normal stress difference ($N_1$) in the channel. The distribution of $N_1$ is shown in three vertical cutting planes in the vicinity of the bubble for $\alpha=0.01$ and $\alpha=0.1$ in Figs. \ref{displacement}f and \ref{displacement}g, respectively, at the same time instant $t^*=5.71$. The higher magnitude of $N_1$ for $\alpha=0.01$ explains why the bubble is pushed towards the channel center. 

 When the elasticity number is increased to $El = 0.1$ by increasing the Weissenberg number in the presence of a strong shear-thinning effect ($\alpha=0.1$) for a spherical bubble ($Ca = 0.01$), the viscoelastic stresses take more time to develop due to a higher relaxation time ($\lambda$) of polymer molecules. Therefore, the enhanced inertia due to the shear-thinning quickly moves the bubble towards the wall while the viscoelastic stresses are not yet fully developed in the flow. The bubble reaches its equilibrium position closer to the wall and its deformation also increases due to a higher shear region there. However, this higher deformability and higher elasticity are not strong enough to reverse the bubble migration back towards the channel center. In the absence of the shear-thinning ($\alpha=0$) with the same high elasticity number, the bubble quickly moves towards the center due to relatively lower inertia.

When the bubble is made more deformable by increasing the capillary number to $Ca=0.1$ (Fig. \ref{displacement}b) in this shear-thinning fluid ($\alpha=0.1$), the deformability-induced lift force resists the pronounced effects of shear-thinning on the fluid inertia. This higher bubble deformation is sufficient enough to push the bubble back toward the channel center. However, due to the resistance by the higher inertial force, the rate of bubble migration is much slower in this fluid than in the non-shear-thinning fluid. When the capillary number is increased even further to $Ca=0.2$, the effect is found to be more pronounced and the rate of bubble migration towards the center of the channel is slightly increased. The enhanced fluid inertia due to the shear-thinning effect also makes the bubble more deformable in a shear-thinning fluid due to an effective high capillary number.

For all the simulations shown in Fig.~\ref{displacement}, the nominal Reynolds number is kept constant. It can be observed that in this non-buoyant situation, the orientation of bubble migration can be controlled with the shear-thinning effect of the viscoelastic fluid. With a strong shear-thinning effect ($\alpha=0.1$), the bubble moves towards the wall and higher elasticity (e.g., $El=0.1$) is not able to reverse its orientation. The higher bubble deformability, however, resists this shear-thinning effect on bubble migration and moves it back towards the center. Interestingly, the bubble deformation increases only by $6\%$ in the case of the maximum capillary number of $Ca=0.2$ in this viscoelastic fluid. However, even this modest deformation makes a significant impact on the dynamics of bubble migration and reverses its orientation towards the center. In the absence of shear-thinning, the bubble migrates towards the center even with a weak elastic effect or with a little deformability. 
\subsection{Path Instability}\label{4.2}
 Path instability of a freely rising bubble in a Newtonian fluid has been thoroughly studied and well documented in the existing literature. For instance, \citet{zenit2008path} experimentally observed that the path instability of a bubble does not depend on the Reynolds number and it is rather governed by the aspect ratio of bubble shape, i.e., its deformation. When the aspect ratio increases beyond $\chi > 2$, vortices generated on the free-slip surface of the bubble give rise to a wake instability behind the trailing edge, forcing its path to become unstable. In the case of a viscoelastic ambient fluid, \citet{shew2006viscoelastic} observed that the flow structure in the wake region becomes more intense leading to path instability even at a much lower deformability.  

 \begin{figure}
  \includegraphics[width=0.98\textwidth]{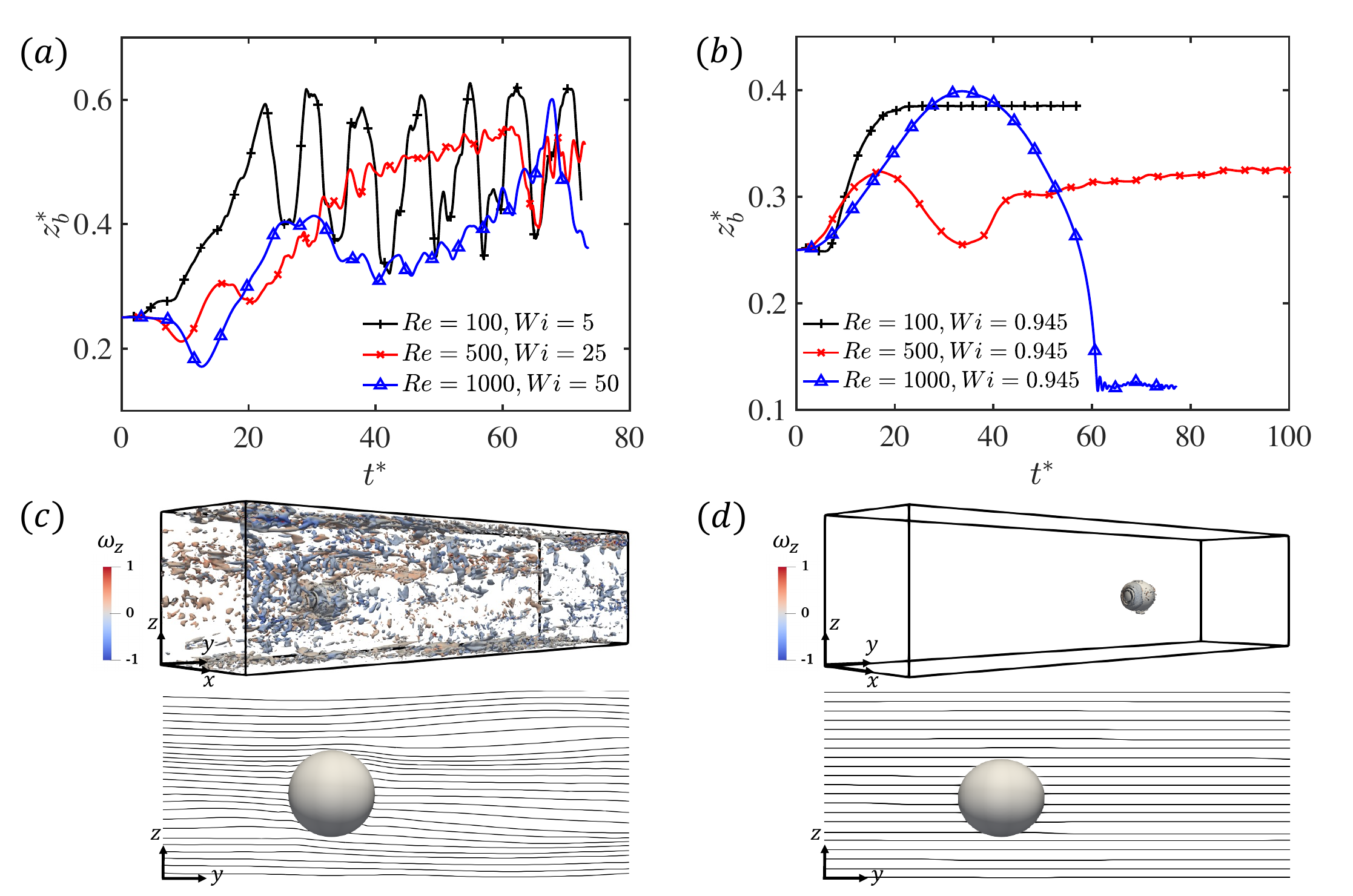}  
  \caption{Evolution of lateral migration of the bubble at constant $El=0.05$ (\textit{a}) and at constant $Wi=0.945$ (\textit{b}) are shown for different values of Reynolds number. (\textit{c}) The constant contours of Q-criterion at $0.001$ colored by the vorticity component ($\omega_z$) are shown for $Re=500, Wi=25$ as the flow becomes elastically unstable due to the curvature of streamlines across the bubble. (\textit{d}) The contours of Q-criterion plotted at the same value of $0.001$ for $Re=500, Wi=0.945$ show a negligible presence, confirming a stable flow. This is also confirmed by the straight streamlines across the bubble.}
 \label{Reynolds}
 \end{figure}

 Simulations are first performed for $Re=100$, $500$, and $1000$ to examine the effects of the Reynolds number on the path instability in the present pressure-driven viscoelastic flow in the absence of buoyancy. To isolate the effects of the Reynolds number, the shear-thinning effect is eliminated by setting $\alpha=0$, and the capillary number is fixed at $Ca=0.01$ to keep the bubble nearly spherical. Simulations are repeated by keeping either the elasticity number or the Weissenberg number constant as the Reynolds number is increased. The results are summarized in Figs.~\ref{Reynolds}a,c and  \ref{Reynolds}b,d for the fixed elasticity number and the fixed Weissenberg number cases, respectively. When the elasticity number is kept constant at $El=0.05$ (Fig. \ref{Reynolds}a), the Weissenberg number increases as the Reynolds number is increased. The bubble path becomes unstable at $Re=100$ (the corresponding $Wi=5$). However, the bubble deformation remains negligible due to a low value of the capillary number. The path instability observed in this case of a nearly spherical bubble at $Re=100$ occurs mainly due to the onset of an elastic flow instability caused by the curved streamlines along the bubble interface. \citet{mckinley1996rheological} determined a critical parameter for the onset of an elastic instability as
\begin{eqnarray}
 M^2=\frac{\lambda U}{R} \frac{N_1}{|\tau_t|} \ge M^2_{crit},
\label{M_crit}
\end{eqnarray} 
\noindent where $U$ is the flow velocity along the streamline, $N_1$ is the first normal stress difference, $R$ is the radius of curvature of the streamline and $\tau_t$ is the total shear stress. \citet{mckinley1996rheological} calculated the critical value for a 2D cylinder as $M_{crit}\approx 6.08$. In the present scenario of a spherical bubble, the critical value turns out to be $M_{crit}\approx 5.59$ at $Re=100$. The constant contours of the Q-criterion (second invariant of velocity gradient tensor) at $0.001$ and the streamlines are also plotted around the bubble in Fig. \ref{Reynolds}c for $Re=500$ to show the overall flow structure. The contours are colored by the magnitude of vorticity in the $z$-direction. The curvature of streamlines across the bubble and the contours of the Q-criterion confirm that the flow is no longer stable. As a result, the bubble path shows an oscillatory pattern around the channel center for $Re=100$. When the Weissenberg number is increased further by increasing the Reynolds number at the fixed value of $El=0.05$, Fig. \ref{Reynolds}a shows that the bubble path becomes more irregular and unpredictable as the flow gets closer to the onset of elastic turbulence stemming from the elastic instability. The exact mechanism behind this probable transition is still elusive (\cite{datta2022perspectives}).

On the other hand, when the Weissenberg number is kept constant at $Wi=0.945$ as $Re$ is increased, the bubble path remains stable even for the Reynolds number as high as $Re=1000$. As shown in Fig.~\ref{Reynolds}d, the streamlines remain straight across the bubble and no significant contours of Q-criterion are observed in the flow, unlike the fixed elasticity number case. It is interesting to observe that the bubble initially moves towards the channel centerline and then reverses its direction towards the channel wall at higher Reynolds numbers. The final position is determined by the interplay of inertial and viscoelastic effects. The equilibrium position of the bubble shifts from the channel center towards the wall as the Reynolds number is increased (Fig. \ref{Reynolds}b). At $Re=1000$, the bubble reaches the wall under the influence of strong inertia, which results in an increase in its deformation as well due to the high shear region near the wall. 


%
\subsection{Effect of Buoyancy}\label{4.3}
The buoyancy is significant in many natural processes and practical applications. Simulations are next performed for a range of E{\"o}tv{\"o}s number to examine the effects of buoyancy while keeping the other parameters fixed at their baseline values of $Re=18.9$, $Ca=0.01$ and $El=0.05$. Note that the Morton number is $M=4.97 \times 10^{-6}$ for $Eo = 1$ in this viscoelastic fluid. For reference, this value is much higher than the Morton number of $2.52 \times 10^{-11}$ for an air bubble in water at $20^\circ$C but can be matched by using a water-glycerin solution (\citet{legendre2012deformation}).

\begin{figure}
  \includegraphics[width=1.0\textwidth]{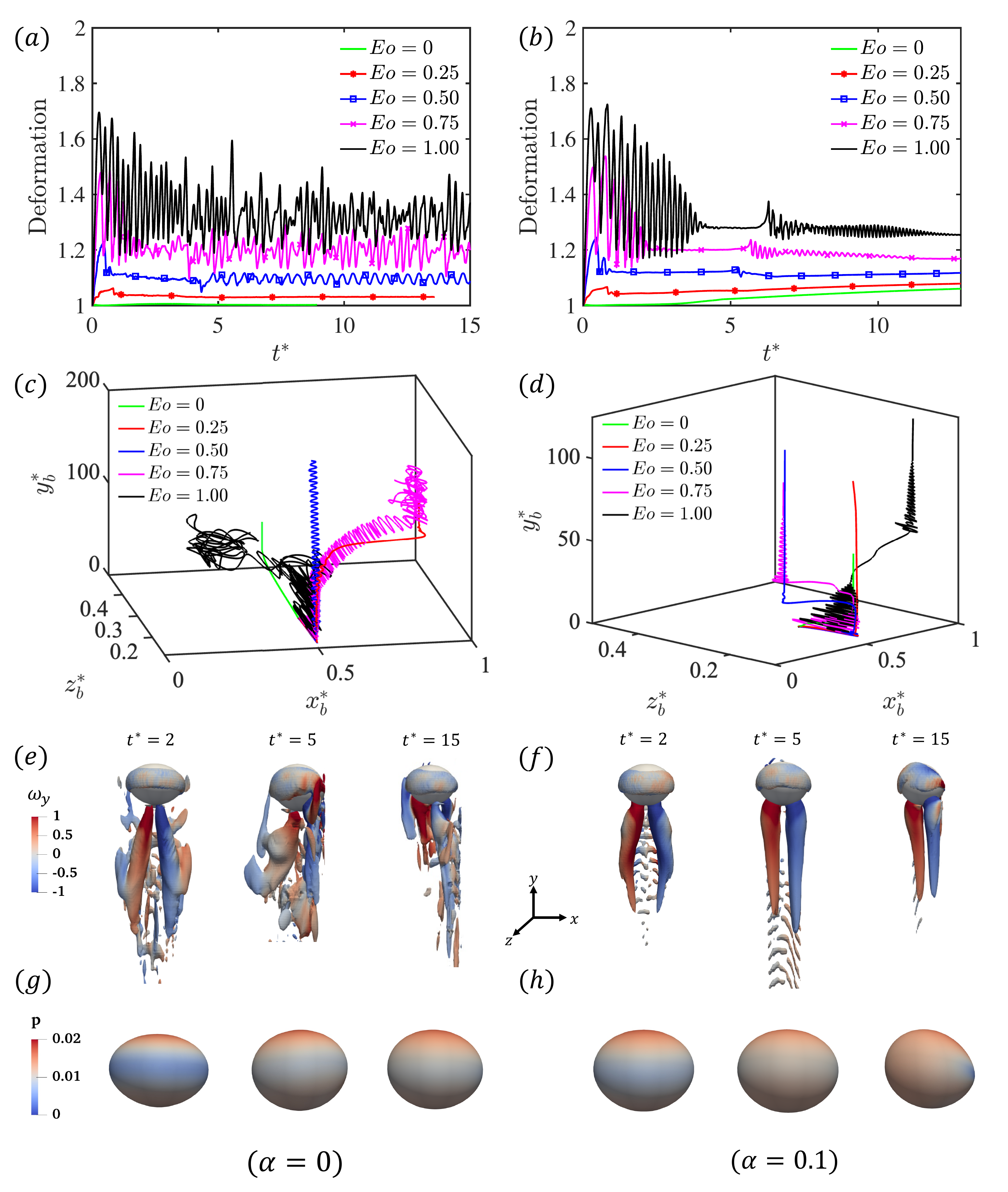}
  \caption{Evolution of bubble deformation and its 3D displacement at different values of E{\"o}tv{\"o}s number are shown for (\textit{a,c}) $\alpha=0$ and for (\textit{b,d}) $\alpha=0.1$. For the $Eo=1$ case, the constant contours of the Q criterion at $0.05$ are shown to visualize the complex flow pattern in the wake region behind the bubble for (\textit{e}) $\alpha=0$ and for (\textit{f}) $\alpha=0.1$. The contours are colored by vorticity ($\omega_y$) in the range of $\pm1$. For the same $Eo=1$ case, the pressure distribution on the bubble surface are shown at the same times for (\textit{g}) $\alpha=0$ and for (\textit{h}) $\alpha=0.1$. ($Re=18.9, Ca=0.01, El=0.05, \beta=0.1$)}
\label{gravity}
\end{figure}

Simulations are performed by gradually increasing the E{\"o}tv{\"o}s number from $Eo=0$ (neutrally buoyant) to $Eo=1$ without and with the shear-thinning effect. The results are shown in Figs.~\ref{gravity}a,c,e,g and \ref{gravity}b,d,f,h in terms of the evolution of bubble deformation, its three-dimensional displacement, the wake structure behind the bubble, and the pressure distribution on the bubble surface  for an Oldroyd-B ($\alpha=0$) and a Giesekus ($\alpha=0.1$) fluid cases, respectively. In the Oldroyd-B fluid,  the bubble moves toward the channel wall without any sign of path instability at $Eo=0.25$. We note that the same bubble moves toward the channel center in the absence of buoyancy. This change in the orientation occurs due to the additional buoyancy-induced lift force (\citet{lu2006dns}). At $Eo=0.5$, the bubble still migrates towards a channel wall (interestingly not to the same wall as in the $Eo=0.25$ case) but small oscillations start to appear in its path with a regular zigzag pattern indicating the onset of a path instability. A similar oscillatory pattern is also visible in the bubble deformation ($\chi$) as seen in Fig.~\ref{gravity}a. When $Eo$ is further increased to $0.75$, the bubble deformation also increases and the bubble path shifts from a zigzag to a helical pattern. The bubble moves laterally to eventually settle near the other side of the channel and its motion becomes more chaotic with a significant velocity component in the wall normal  ($x$) direction. At $Eo=1$, the bubble path becomes completely unstable with a deformation parameter ($\chi$) exceeding $1.6$ (Fig. \ref{gravity}a). 

It is worth noting that the deformation parameter is only about $\chi=1.1$ when the bubble undergoes path instability at $Eo=0.5$ in the present setting. This value is much smaller than the condition $\chi\ge 2$ required for the onset of a path instability in the case of a freely rising bubble in a Newtonian fluid (\citet{zenit2008path}). This result confirms that the viscoelasticity facilitates the earlier onset of path instability as also pointed out by \citet{shew2006viscoelastic}. The wake region is much more intense in the viscoelastic fluid than in the Newtonian fluid due to the additional viscoelastic stresses, forcing its path to become unstable at a relatively small deformation. The evolution of the wake region and the corresponding pressure distribution on the bubble surface is depicted for the $Eo=1$ case in Figs.~\ref{gravity}e and \ref{gravity}g, respectively, to qualitatively show the mechanism for the path instability. As seen, the structure of the wake is highly complex and transient, and the pressure distribution changes accordingly, indicating the footprint of the instability. 

Next, a strong shear-thinning effect is added ($\alpha=0.1$), and the simulations are repeated for the same values of $Eo$ by keeping the other parameters fixed at their baseline values of $Re=18.9$, $El=0.05$, and $Ca=0.01$. The results are summarized on the right side of Fig.~\ref{gravity}. As seen in Fig.~\ref{gravity}d, the bubble migrates towards a wall for all the cases but interestingly not to the same wall. The bubble path remains stable for $Eo=0.25$ but it migrates towards the wall at a much higher rate than that for the non-buoyant case of $Eo=0$. This behavior is attributed to the same enhanced lift force due to the buoyancy as in the non-shear-thinning case. Some oscillations appear in the bubble trajectory for $Eo=0.5$ but they are damped very quickly as the bubble stabilizes near a wall. As $Eo$ is increased beyond $Eo=0.5$, the oscillations are amplified leading to path instability. For $Eo=0.75$, the bubble migrates laterally and settles near the other wall similar to the Oldroyd-B fluid case. In this case, the bubble initially follows a helical path but, interestingly, its path is stabilized after some time (Fig. $\ref{gravity}d$). The corresponding oscillatory pattern observed in the deformation also vanishes once the bubble path is stabilized (Fig. $\ref{gravity}b$). The stabilized bubble path is also observed for $Eo=1$. The stabilization mechanism is visualized in Fig.~\ref{gravity}f where the evolution of the wake structure is depicted for the $Eo=1$ case. As seen, the wake behind the bubble is not symmetric initially but it later evolves into a symmetric two counter-rotating vortex structures indicating that the path is stabilized. Two contributory factors are speculated to be the main reason for this stabilization mechanism. First, with a decrease in fluid viscosity in the vicinity of the bubble due to the shear-thinning effect, the vortices generated on the surface of the deformed bubble are not strong enough to make the wake region unstable. Secondly, due to the shear-thinning of the fluid, the first normal stress difference $N_1$ decreases in this viscoelastic fluid, which reduces the destabilization effect of viscoelasticity. Thus, the intense flow structures observed in the wake region in the case of the Oldroyd-B fluid (Fig.~\ref{gravity}e) are damped out in this shear-thinning fluid (Fig.~\ref{gravity}f). Although an addition of shear-thinning makes the bubble path more stable, it cannot maintain path stability indefinitely. For instance, although not shown here due to space consideration, once the bubble deformation exceeds $\chi>2$ at $Eo=2$, the path remains unstable even for the mobility factor as high as $\alpha=0.1$. 

\begin{figure}
  \includegraphics[width=1.0\textwidth]{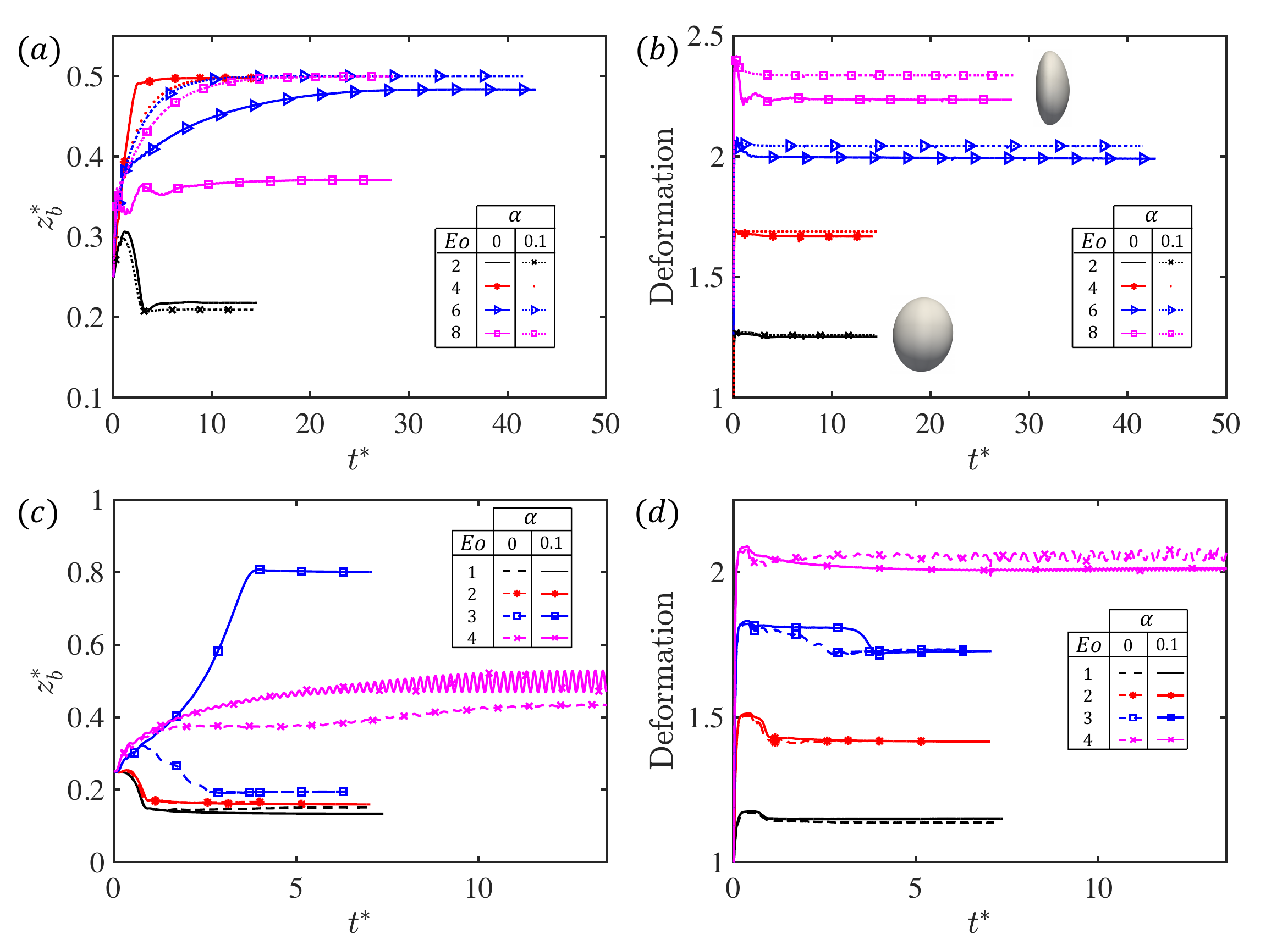}
  \caption{(\textit{a}) Evolution of bubble displacement in the lateral ($z$) direction and (\textit{b}) its deformation for different values of Eötvös number. For (\textit{a}) and (\textit{b}), $Re=18.9,\; Ca=0.01, \; El=0.05, \; \beta=0.9$. For (\textit{c}) and (\textit{d}), $Re=47.25, \; Ca=0.01, \; El=0.02, \; \beta=0.9$. The bubble shapes are shown in the inset of (\textit{b}) for $\alpha=0$ case at $Eo=2$ and $Eo=8$.}
\label{beta}
\end{figure}

All the simulations have heretofore been performed for a small value of the viscosity ratio, $\beta=0.1$, representing highly concentrated polymer solutions. Thus, the fluid exhibited a strong viscoelastic behavior and the bubble path remained stable only up to $Eo=0.25$ and $Eo=1$ without and with the shear-thinning effect, respectively. To examine the bubble migration dynamics in dilute polymer solutions, simulations are now performed for another extreme value of $\beta = 0.9$ without and with the shear-thinning effect. Figures~\ref{beta}a and \ref{beta}b show the evolution of lateral displacement of a bubble and its deformation for the range of $2\le Eo \le 8$. We note that, at the lower values of $Eo$ up to $Eo=2$, there is not much difference observed in the bubble deformation and its equilibrium position compared to the case of $Eo=2$ in Fig.~\ref{beta}a. As $Eo$ increases, the difference in the equilibrium position of the bubble becomes more and more pronounced without and with the shear-thinning effect. For the smaller values of E{\"o}tv{\"o}s number up to $Eo=2$, the shear-thinning effect does not play a dominant role due to a lower value of bubble deformation. As the contribution of polymeric viscosity towards the total viscosity of the fluid is already very low at $\beta=0.9$, a further decrease in the viscosity due to shear-thinning starts to show its effect only at a higher value of bubble deformation. Beyond $Eo=2$ and with a strong shear-thinning ($\alpha=0.1$), the bubble reaches the channel center for all the values of $Eo$ up to $Eo=8$, whereas, without the shear-thinning effect, the equilibrium position of the bubbles occurs slightly away from the center depending upon its deformation. The most interesting feature is the persistent stability of the bubble path even when the deformation exceeds $2.3$ for the $Eo=8$ case (Fig. \ref{beta}b-inset). \citet{zenit2008path} have shown that for a freely rising bubble, the path becomes unstable when the deformation value exceeds $2$. They argued that for such a buoyancy-driven bubble, path instability depends upon the bubble deformation rather than the Reynolds number. On the other hand, \citet{haberman1953experimental} and \citet{kelley1997path} had earlier demonstrated that the path instability of an air bubble in a viscous fluid of a Hele-Shaw cell did depend on the Reynolds number. It is found that the criterion, $\chi > 2$, for triggering path instability does not apply in the present case of a pressure-driven viscoelastic channel flow at this low Reynolds number.

In order to verify the dependence of bubble path instability on the Reynolds number, simulations are next performed by increasing the Reynolds number to $47.25$ while keeping the Weissenberg number fixed at $Wi=0.945$. As a result, the elasticity number is reduced to $0.02$. Simulations are repeated for $\alpha = 0$ (no shear-thinning) and $\alpha = 0.1$ (significant shear-thinning) cases for the range of E{\"o}tv{\"o}s number $1\le Eo \le 4$.  Figures~\ref{beta}c and \ref{beta}d depict the evolution of bubble displacement in the lateral direction and its deformation. As $Eo$ increases, the bubble deformation also increases and its equilibrium position gradually shifts away from the wall. At $Eo=3$, as the fluid viscosity decreases in the vicinity of the bubble, the elastic effect becomes dominant and the bubble migrates towards the channel center. The bubble crosses the channel center under the combined effects of deformation-induced lift force and the fluid elasticity and settles at an equilibrium position closer to the wall on the other side of the channel as the inertial effects eventually dominate (Fig. \ref{beta}c). On the other hand, without shear-thinning, the bubble migrates towards the wall as the deformation and elasticity are not strong enough to counter high inertial lift force at $Eo=3$. Interestingly, both the bubbles settle at the same equilibrium position closer to the wall but on the opposite sides of the channel.

At $Eo=4$, when the bubble deformation exceeds $2$ and the shear-thinning effect is present, small oscillations start to appear in the bubble path indicating the onset of a path instability. It shows that path instability of the bubble in this pressure-driven viscoelastic flow is a function of Reynolds number as well along with the bubble deformation. It is also observed that the bubble path remains stable in the absence of shear-thinning ($\alpha=0$) at the same value of $Eo=4$. Interestingly, this role of shear-thinning is opposite to what was observed in the case of $\beta = 0.1$, i.e., the shear-thinning damps out the path instability at low values of $\beta$ (high polymer concentration) whereas it promotes the path instability of the bubble at high $\beta$ values (dilute polymer concentration). This interesting behavior is attributed to the local change in the effective elasticity number ($El$) due to the shear-thinning effect. \citet{ray2014absolute} have shown that, for larger values of $El$, there is an asymptotic limit of stabilization observed in the Oldroyd-B model. We find that this role of elasticity number in stabilizing or destabilizing the bubble path is further dependent upon the value of $\beta$. At lower $\beta$, a decrease in the local value of $El$ stabilizes the bubble path while, at the higher value of $\beta$, the same decrease in $El$ is found to be destabilizing.
  
\subsection{Bubble-Induced Secondary Flow}\label{4.4}

\begin{figure}
  \includegraphics[width=1.0\textwidth]{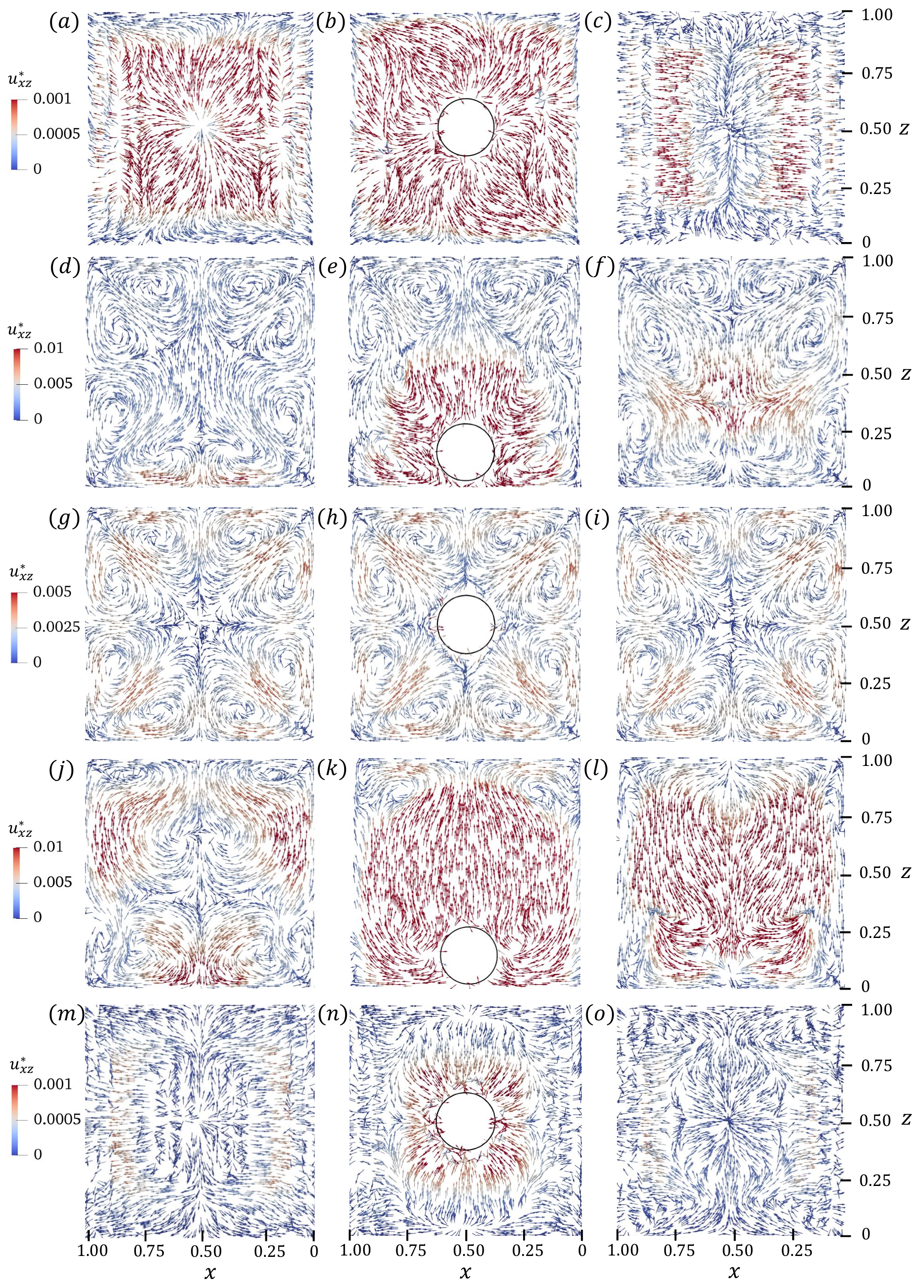}
  \caption{Velocity vectors representing the complex flow patterns of secondary flow field in vertical cutting planes at a downstream (\textit{a,d,g,j,m}), the center (\textit{b,e,h,k,n}) and an upstream (\textit{c,f,i,l,o}) of the bubble. (\textit{a,b,c}) $Re=18.9, Ca=0.01, El=0.05, \alpha=0$. (\textit{d,e,f}) $Re=18.9, Ca=0.01, El=0.05, \alpha=0.1$. (\textit{g,h,i}) $Re=18.9, Ca=0.2, El=0.05, \alpha=0.1$. (\textit{j,k,l}) $Re=18.9, Ca=0.01, El=0.1, \alpha=0.1$. (\textit{m,n,o}) $Re=18.9, Ca=0.1, El=0.05, \alpha=0$.}
\label{sec_flow}
\end{figure}

The presence of the second normal stress difference ($N_2$) and the geometry of the square-shaped channel induces a secondary flow field in a direction perpendicular to the primary flow. The dynamics of this secondary flow field are significantly influenced by the presence of a bubble. Moreover, the orientation of bubble migration also affects the magnitude of the secondary flow velocity. The secondary flow velocity has been analyzed by varying the bubble deformability ($Ca$), the shear-thinning ($\alpha$), the fluid elasticity ($Wi$), and the fluid inertia ($Re$). Figure~\ref{sec_flow} shows complex flow patterns in vertical cutting planes at the downstream, the center, and upstream of the bubble under different flow conditions. The upstream and downstream planes are located at $+0.75$ and $-0.75$ from the bubble center, respectively. The velocity vectors are colored by the magnitude of flow velocity in the $xz-$plane. In an Oldroyd-B fluid, the particular distribution of $N_1$ in the channel far away from the bubble builds up a weak secondary flow oriented towards the corners of the channel from its center. It generates a `flower-shaped' flow pattern as shown in Fig.~\ref{sec_flow}a. The presence of $N_2$ due to the shear-thinning effect (Giesekus fluid) generates an additional eight vortices at the corners of the channel (Fig. \ref{sec_flow}d). Once the bubble passes through this plane and migrates towards the center (Fig.~\ref{sec_flow}g and \ref{sec_flow}h), this flow pattern gets disturbed but becomes symmetric again above and below the bubble once the bubble attains its steady state position. On the other hand, if the bubble migrates towards a channel wall, the flow pattern becomes highly asymmetric away from the wall resulting in a strong secondary flow with a velocity as high as the order of bubble migration velocity (Fig.~\ref{sec_flow}k).      

\begin{figure}
  \includegraphics[width=1.0\textwidth]{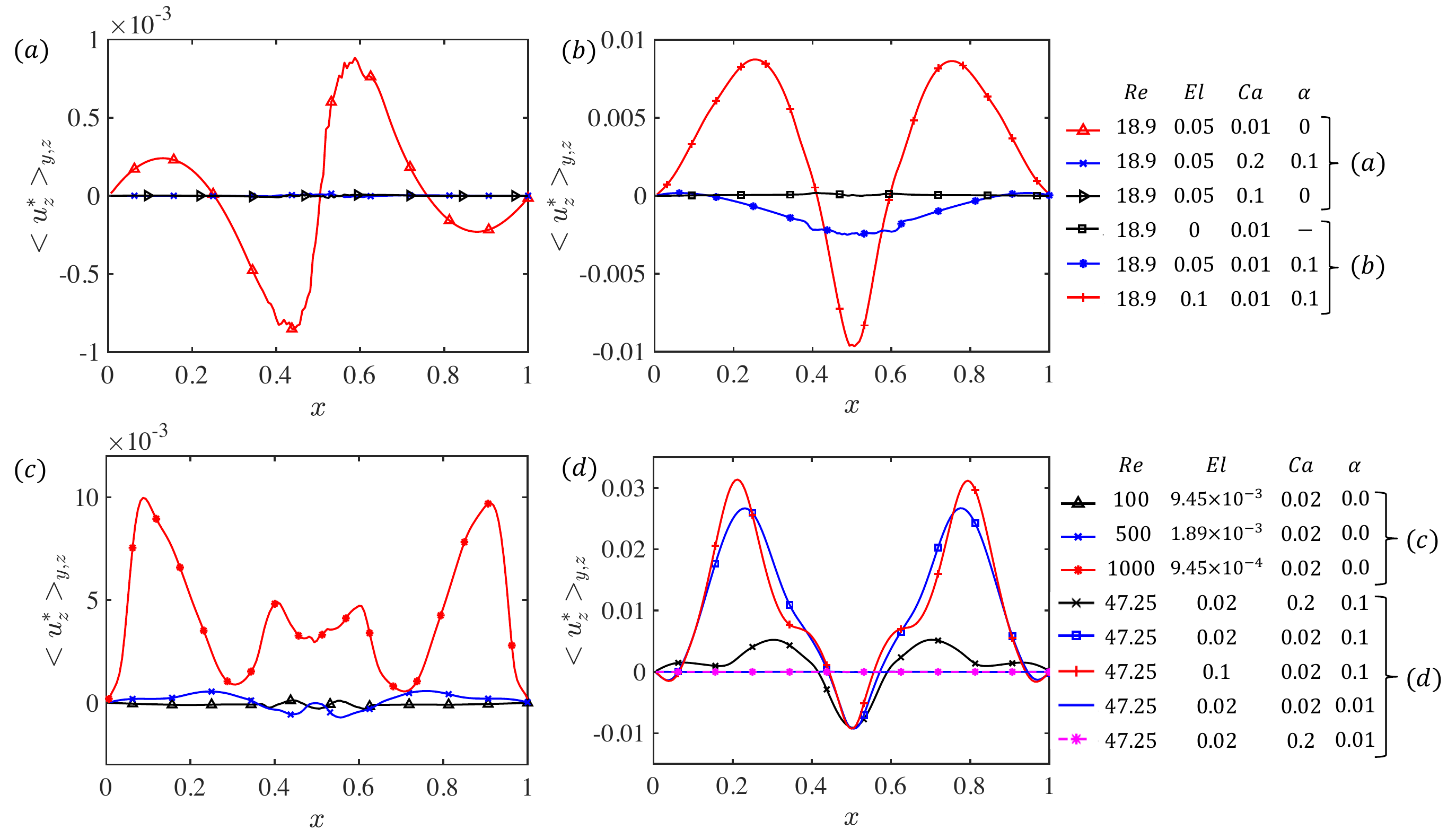}
  \caption{The averaged secondary flow velocity under the flow conditions (\textit{a}) where the bubble migrates towards the channel center and (\textit{b}) where the bubble migrates towards a wall at $Re=18.9$. The averaged secondary flow velocity is plotted for higher Reynolds numbers in the range $47.25\leq Re \leq 1000$ for (\textit{c}) an Oldroyd-B fluid and (\textit{d}) a Giesekus fluid.}
\label{secflow_plots}
\end{figure}

The quantitative comparison of secondary flow velocities is summarized in Fig.~\ref{secflow_plots} for different flow conditions. The $z$-component of flow velocity is numerically integrated in the region $\pm0.75$ upstream and downstream of the bubble in the streamwise ($y$) direction and the integration is performed in the entire domain [$0 , 1$] in the $z$-direction, i.e., 

\begin{eqnarray}
\langle u_z\rangle_{y,z}=\frac{1}{A_{yz}}\int_0^1\int_{-0.75}^{+0.75}u_z dzdy 
\cong 
\frac{1}{\left(j_{\rm max}-j_{\rm min}+1\right) N_z}\sum_{k=1}^{N_z}\sum_{j=j_{\rm min}}^{j_{\rm max}}(u_z)_{jk},
\label{sec_vel}
\end{eqnarray}

 \noindent where $A_{yz}$ is the area of the $yz$-plane, and $j_{\rm min} = (y_b-0.75)N_y/L$ and $j_{\rm max} = (y_b+0.75)N_y/L$ with $N_y$ and $N_z$ being the number of grid points in the $y$ and $z$ directions, respectively.  The average secondary velocity, $\langle u_z \rangle_{y,z}$, in the $y$ and $z$ directions is plotted along $x$ in Fig.~\ref{secflow_plots} once the bubble reaches its equilibrium position closer to a wall or towards the center of the channel. Figures~\ref{secflow_plots}a and \ref{secflow_plots}b show the average secondary flow velocity for $Re=18.9$ while the results for higher Reynolds numbers are shown in Figs.~\ref{secflow_plots}c and \ref{secflow_plots}d. 

At $Re=18.9$, the magnitude of the secondary flow velocity remains negligible for all the situations where the bubble migrates towards the channel center (Fig.~\ref{secflow_plots}a). In an Oldroyd-B fluid, the bubble migrates towards the center of the channel under the effect of elasticity, as shown in Fig.~\ref{displacement}. As the bubble migrates in the $z$-direction and reaches the center, the flow pattern becomes symmetric above and below the bubble (Fig.~\ref{sec_flow}b). In a Giesekus fluid, when the shear-thinning effect is sufficient enough to move the bubble towards a wall ($\alpha=0.1$), a strong secondary flow pattern is observed as shown in Fig.~\ref{secflow_plots}b. When the elasticity number is increased in this shear-thinning fluid, the maximum magnitude of the secondary flow velocity is attained. This is the same combination of parameters for which the bubble equilibrium position is closest to the channel wall in Fig.~\ref{displacement}a. The flow pattern above the bubble for this situation is depicted in Fig.~\ref{sec_flow}k.
When the Reynolds number is increased in an Oldroyd-B fluid, the equilibrium position of the bubble moves closer to the channel wall. When the Weissenberg number is kept constant while increasing the Reynolds number to avoid elastic instability, as discussed in Section \ref{4.2}, the magnitude of the secondary flow velocity keeps increasing with the Reynolds number and becomes as high as $1 \%$ of the maximum flow velocity in the channel (Fig. \ref{secflow_plots}c). A small asymmetry observed in the secondary flow pattern for the $Re=1000$ case is attributed to the slight migration of bubble in the $x$-direction due to onset of an inertial path instability.

At $Re=47.25$, when the shear-thinning effect is small ($\alpha = 0.01$), the bubble moves towards the center of the channel whether it is spherical ($Ca=0.01$) or highly deformable ($Ca=0.1$). The secondary flow velocity remains negligible for both of these situations (Fig. \ref{secflow_plots}d). However, this time the magnitude of the secondary flow velocity depends on the bubble deformability along with the fluid elasticity. When the bubble deformability is increased, the magnitude of the secondary flow velocity is reduced whereas an increase in the fluid elasticity ($Wi$) further increases its magnitude. Thus, a nearly spherical bubble migrating towards the channel wall induces maximum secondary flow due to the maximum asymmetric area of influence above. This area reduces with the bubble deformability and hence a reduction in the induced secondary flow is observed. Similarly, a higher magnitude of viscoelastic stresses due to higher $Wi$ generates a complex flow pattern in the direction normal to the primary flow. This flow pattern becomes more and more asymmetric with increasing $Wi$ and as a result, the secondary flow velocity becomes as high as $3\%$ of the maximum flow velocity in this square channel (Fig. \ref{secflow_plots}d). 

It can be summarized that an increase in the fluid elasticity ($Wi$) or inertia ($Re$) increases the magnitude of the secondary flow velocity as long as the bubble moves towards a wall. If the flow conditions are such that the bubble moves towards the channel center, an increase in the inertia or elasticity of the fluid makes negligible impact on the velocity of the secondary flow once the bubble reaches its steady state position. However, during the transient period while the bubble is still migrating towards the center, the inertia and fluid elasticity do make an impact on the velocity of the secondary flow but its magnitude during this transient period remains two orders of magnitude lower than the corresponding cases where the bubble migrates towards the wall.   

It is important to note that the secondary flow velocity induced by a bubble is at least an order of magnitude higher than the velocity induced by the solid particle under the similar flow conditions as reported by \cite{li2015dynamics}. The primary reason is the higher drag experienced by the solid particle due to the no-slip condition on its surface. This attribute of bubble-induced secondary flow can be exploited in the situations where mixing, heat transfer or other transport phenomena are of importance.
\subsection{Flow Start-up}\label{4.5}
It is a well-known aspect that stress waves propagate (\cite{duarte2008numerical}) and cause velocity fluctuations during flow build-up in a channel flow. The elasticity of the fluid affects these fluctuations depending upon the relaxation time of its polymer molecules. In our earlier work (\citet{naseer2023dynamics}), these velocity fluctuations were also observed in the rise velocities of freely rising bubbles in a viscoelastic fluid. Similarly, these velocity fluctuations affect the migration velocity of the solid particle as well during the flow build-up (\cite{rajagopalan1996sedimentation}). \cite{goyal2012direct} have shown that a solid particle settling down during these fluctuations often rebounds after the first oscillation. In the present study, we explore the same effects for a deformable bubble in this viscoelastic channel flow.

\begin{figure}
  \includegraphics[width=1.0\textwidth]{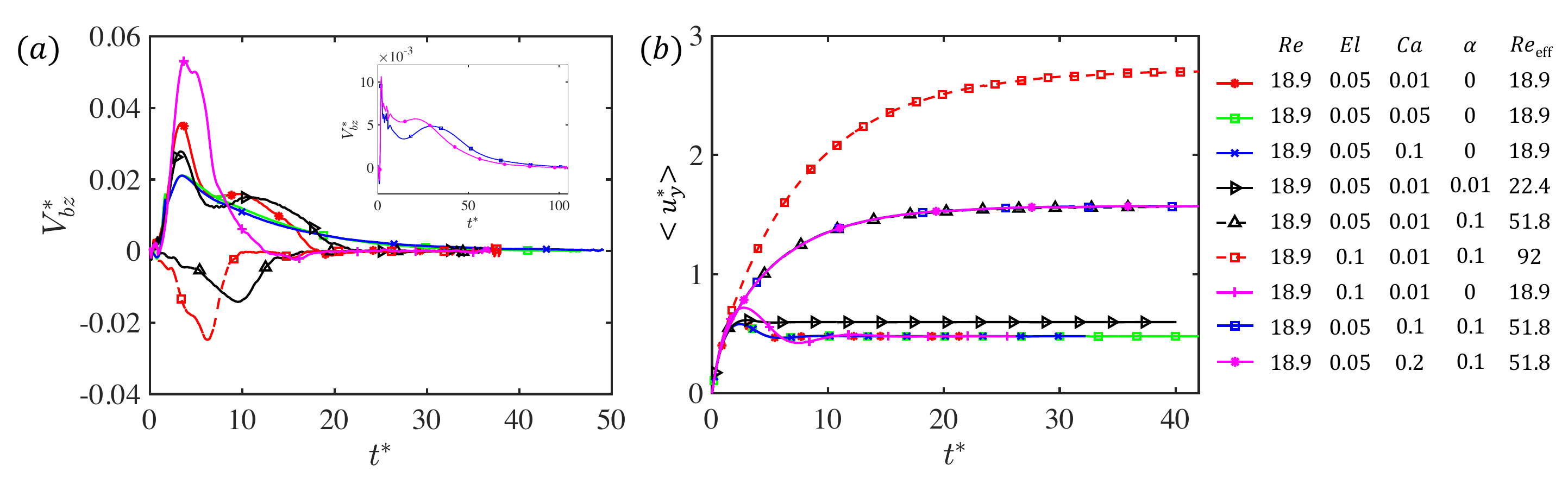}
  \caption{(\textit{a}) Evolution of bubble migration velocity in the wall normal ($z$) direction during flow start-up. (\textit{b}) Evolution of average steamwise component of flow velocity in the channel.}
\label{migration_velocity}
\end{figure}

Figure~\ref{migration_velocity}a shows the migration velocity of the bubble in the wall-normal direction  ($z$) under various flow conditions. The two cases are shown in the inset where the bubble is highly deformable ($Ca=0.1$ and $Ca=0.2$) and it takes a comparatively longer time to reach a steady state. Figure~\ref{migration_velocity}b shows the evolution of the average streamwise component of flow velocity in the channel for the same nine situations. In the case of a nearly spherical bubble (low $Ca$) migrating in an Oldroyd-B fluid ($\alpha = 0$), the elastic stresses develop quickly in the flow and take the bubble towards the center. As $N_1$ decreases in the low shear region near the center, the bubble reaches its peak velocity at $t^* \approx 4$ and then starts to decelerate while continuing its migration towards the center. The minor fluctuations in the bubble velocity are attributed to the extension/relaxation of polymer molecules across its surface as recently explained by \cite{bothe2022molecular} and verified by \cite{naseer2023dynamics}. A minor decline in the flow velocity is observed after an initial peak once the viscoelastic stresses are developed in the flow (Fig. \ref{migration_velocity}b). When the bubble deformability is increased ($Ca=0.05$ and $Ca=0.1$), the bubble starts migrating towards the channel center. However, due to a slight increase in its deformability, the peak velocity of the bubble migration remains lower than that of the nearly spherical bubble.
When a spherical bubble ($Ca=0.01$) migrates in the Giesekus fluid having a very low shear-thinning ($\alpha=0.01$), migration towards the channel center occurs primarily due to the elastic effects and similar fluctuations  are observed in the migration velocity as in the case of Oldroyd-B fluid. As the shear-thinning promotes the relative importance of fluid inertia which acts to pull the bubble back towards the wall, the peak velocity of the bubble remains lower than that in the Oldroyd-B fluid. The average flow velocity is slightly increased in the channel due to the shear-thinning effect and the actual effective Reynolds number becomes $Re_{\rm eff}=22.4$. When the shear-thinning effect is high enough ($\alpha = 0.1$), the enhanced fluid inertia overcomes the elastic effects and causes the bubble to migrate towards the wall, as indicated by the opposite direction of bubble velocity in Fig. \ref{migration_velocity}a. The flow velocity in the channel increases further and the effective Reynolds number reaches $51.8$ with the same value of an applied pressure gradient. When the elasticity number is increased ($El=0.1$) by increasing the polymer relaxation time in this shear-thinning viscoelastic fluid, the inertial effects still dominate and push the bubble towards a wall at a higher rate. With increased viscoelastic stresses at higher $El$ and a strong shear-thinning effect, the effective Reynolds number reaches $92$. With the same high elasticity number ($El=0.1$) but without shear-thinning ($\alpha = 0$), the bubble migrates towards the center under the elastic effects. Due to a higher Weissenberg number, the viscoelastic stresses become comparatively stronger and the bubble reaches a higher value of its migration velocity. As there is no shear-thinning effect, the flow Reynolds number remains at the baseline value of $Re=18.9$.

Once the bubble is made highly deformable ($Ca=0.1$ and $Ca=0.2$) and the fluid is strongly shear-thinning ($\alpha=0.1$), the `deformability-induced' lift force overcomes the shear-thinning enhanced inertial effects and the bubble moves towards the channel center. As the bubble gets closer to the center, its deformability starts reducing due to low shear (as shown in Fig.~\ref{displacement}b) resulting a deceleration in its lateral migration velocity. It is worth noting that the effective flow Reynolds number becomes $Re_{\rm eff} = 51.8$ due to a strong shear-thinning effect in this case.   

\section{Conclusions}\label{Conclusion}

 Interface-resolved numerical simulations are performed to explore the complex dynamics of cross-stream migration of a bubble in a viscoelastic pressure-driven channel flow for a wide range of flow parameters. Unlike the solid particle, the bubble deformability, the slip condition at the interface, and  the bubble-induced elastic flow instability at a higher Weissenberg number add further complexity to this phenomenon. The bubble deformability ($Ca$), shear-thinning of the ambient fluid ($\alpha$), fluid elasticity ($Wi$), buoyancy ($Eo$) and concentration of polymers ($\beta$) are varied one by one to isolate their sole effects. The conditions for triggering a path instability of a bubble in this pressure-driven viscoelastic flow are investigated and the bubble-induced secondary flow is quantified.
 
 It is observed that the forces induced by fluid inertia push the bubble towards the wall while elasticity and deformability pull it towards the channel center. This interplay between fluid inertia, elasticity, and bubble deformability determines the orientation of bubble migration in the channel and its final equilibrium position. As shear-thinning increases the relative importance of the fluid inertia, it is found to promote migration towards a wall. At a high Weissenberg number, elastic instability occurs due to the curvature of streamlines across the bubble. When the Reynolds number is increased while keeping the Weissenberg number fixed at a low value, the bubble path remains stable even for the Reynolds number as high as $Re=1000$.

In the case of a high polymer concentration (e.g., $\beta=0.1$), it is found that strong viscoelastic effects make the bubble path unstable even when the bubble is nearly spherical. This path instability occurs as the flow itself becomes elastically unstable at a higher value of $Wi$. In the absence of strong viscoelastic stresses (low $Wi$), the path of the bubble becomes unstable when its deformation becomes high due to high $Eo$. However, the shear-thinning acts as a stabilizing factor and can make the bubble path stable even for the bubble deformation as large as $\chi = 2$. The stabilizing role of shear-thinning is reversed at higher values of $\beta$. When the Reynolds number is small and $\beta=0.9$, the bubble path remains stable even for the E{\"o}tv{\"o}s number as large as $Eo=8$ and its deformation exceeds $\chi > 2.3$. Thus, the path instability of a bubble in a viscoelastic pressure-driven channel flow can occur when the flow itself becomes elastically unstable at high $Wi$ or when its deformation exceeds a threshold value under buoyant conditions. If the path instability is triggered by the later mechanism, i.e., when the bubble deformation exceeds a critical value, the shear-thinning suppresses the path instability at higher polymer concentration (lower $\beta$) while it reverses its role and promotes the path instability at lower polymer concentration (higher $\beta$). Furthermore, it is found that the threshold value of bubble deformation to trigger path instability is a function of Reynolds number as well in this pressure-driven viscoelastic flow unlike the path instability of a freely rising bubble in a Newtonian flow (\citet{zenit2008path}).

It is shown that the secondary flow velocity induced by a bubble migrating towards the wall is an order of magnitude higher than the one induced by a solid particle under similar flow conditions mainly due to the free slip condition on its surface. For all the situations where the bubble migrates towards the center, the secondary flow velocity becomes negligible due to the symmetric flow pattern above and below the bubble when viewed in its migrating plane.

During the flow build-up in this square-shaped channel, the maximum value of flow velocity depends upon the shear-thinning property of the fluid. The bubble migration velocity is dictated by the inertial effects and the rate at which the viscoelastic stresses are built up in the flow. 

  
\section{Acknowledgment}

We acknowledge financial support from the Scientific and Technical Research Council of Turkey (TUBITAK) [Grant Numb 119M513] and the Research Council of Finland [Grant Number 354620].

\section{Data availability and supplementary material}

Data can be made available upon request. A time-lapse movie showing the generation and growth of instabilities in the flow at a high Weissenberg number ($Re=500, Wi=25$) due to the presence of a bubble can be accessed in the supplementary material.

\begin{center}
\section*{Appendix}
\end{center}

\begin{figure}
  \includegraphics[width=1.0\textwidth]{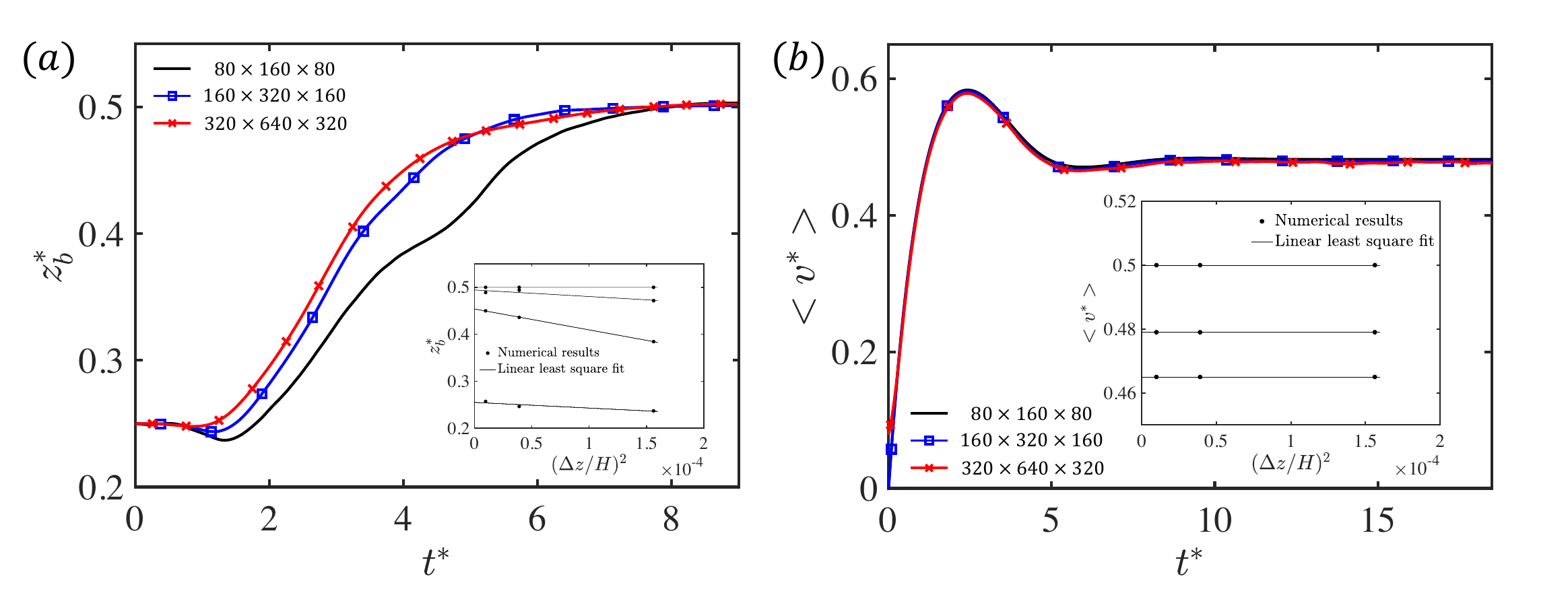}
  \caption{(\textit{a}) Evolution of bubble displacement in the wall-normal ($z$) direction and (\textit{b}) the average streamwise component of flow velocity in the channel are shown at different grid resolutions. Linear least-squares fits at different time intervals are shown in the respective insets. ($Re=18.9, El=0.05, Ca=0.01, \alpha=0$)}
\label{grid}
\end{figure}

\noindent A comprehensive grid convergence study is performed to ascertain the numerical accuracy of the results reported in this study. Figure \ref{grid}a shows the evolution of bubble displacement in the wall-normal ($z$) direction plotted at three different grid resolutions. Linear least-squares fits at $t^* = 1.38, 4, 6$ and $8$ are shown in the inset of Fig.~\ref{grid}a where the approximately linear relations confirm the expected second-order spatial accuracy of the numerical scheme. These results show that a grid resolution containing $160 \times 320 \times 160$ grid cells in the $x$, $y$, and $z$ directions, respectively, is sufficient to reduce the spatial error below $3.5\%$. Therefore this grid resolution is used in all the simulations reported in this paper. Note that, for this grid resolution, there are about $40$ grid points in the $x$ and $z$ directions and $20$ grid points in the streamwise ($y$) direction per bubble diameter. Figure~\ref{grid}b shows the evolution of average flow velocity in the channel and the linear least-squares fits shown in its inset confirm that the numerical error remains negligible for this quantity even with a grid size of $80 \times 160 \times 80$. The time step size is strictly restricted by the stability constraints in the present explicit numerical method and the temporal error remains negligible compared to the spatial error (\cite{tryggvason2001front,tryggvason2011direct}). Therefore, a temporal error convergence is not examined.

\bibliographystyle{jfm}
\bibliography{References}
\end{document}